\documentstyle[12pt,epsf,epsfig]{article}

\catcode`\@=11
\long\def\@makefntext#1{
\protect\noindent \hbox to 3.2pt {\hskip-.9pt  
$^{{\ninerm\@thefnmark}}$\hfil}#1\hfill}		

\def\@makefnmark{\hbox to 0pt{$^{\@thefnmark}$\hss}}  
	
\def\ps@myheadings{\let\@mkboth\@gobbletwo
\def\@oddhead{\hbox{}
\rightmark\hfil\ninerm\thepage}   
\def\@oddfoot{}\def\@evenhead{\ninerm\thepage\hfil
\leftmark\hbox{}}\def\@evenfoot{}
\def\sectionmark##1{}\def\subsectionmark##1{}}

\renewcommand{\thefootnote}{\fnsymbol{footnote}}

\newcounter{sectionc}\newcounter{subsectionc}\newcounter{subsubsectionc}
\renewcommand{\section}[1] {\vspace*{0.6cm}\addtocounter{sectionc}{1} 
\setcounter{subsectionc}{0}\setcounter{subsubsectionc}{0}\noindent 
	{\normalsize\bf\thesectionc. #1}\par\vspace*{0.4cm}}
\renewcommand{\subsection}[1] {\vspace*{0.6cm}\addtocounter{subsectionc}{1} 
	\setcounter{subsubsectionc}{0}\noindent 
	{\normalsize\it\thesectionc.\thesubsectionc. #1}\par\vspace*{0.4cm}}
\renewcommand{\subsubsection}[1]
{\vspace*{0.6cm}\addtocounter{subsubsectionc}{1}
	\noindent {\normalsize\rm\thesectionc.\thesubsectionc.\thesubsubsectionc. 
	#1}\par\vspace*{0.4cm}}

\newcounter{appendixc}
\newcounter{subappendixc}[appendixc]
\newcounter{subsubappendixc}[subappendixc]

\renewcommand{\appendix}[1] {\vspace*{0.6cm}
        \refstepcounter{appendixc}
        \setcounter{figure}{0}
        \setcounter{table}{0}
        \setcounter{equation}{0}
        \renewcommand{\thefigure}{\Alph{appendixc}.\arabic{figure}}
        \renewcommand{\thetable}{\Alph{appendixc}.\arabic{table}}
        \renewcommand{\theappendixc}{\Alph{appendixc}}
        \renewcommand{\theequation}{\Alph{appendixc}.\arabic{equation}}
        \noindent{\bf Appendix \theappendixc #1}\par\vspace*{0.4cm}}

\def\abstracts#1{{
	\centering{\begin{minipage}{12.2truecm}\footnotesize\baselineskip=12pt\noindent
	\centerline{\footnotesize ABSTRACT}\vspace*{0.3cm}
	\parindent=0pt #1
	\end{minipage}}\par}} 

\renewenvironment{thebibliography}[1]
	{\begin{list}{\arabic{enumi}.}
	{\usecounter{enumi}\setlength{\parsep}{0pt}
\setlength{\leftmargin 1.25cm}{\rightmargin 0pt}
	 \setlength{\itemsep}{0pt} \settowidth
	{\labelwidth}{#1.}\sloppy}}{\end{list}}

\newcounter{itemlistc}
\newcounter{romanlistc}
\newcounter{alphlistc}
\newcounter{arabiclistc}

\newcommand{\fcaption}[1]{
        \refstepcounter{figure}
        \setbox\@tempboxa = \hbox{\footnotesize Fig.~\thefigure. #1}
        \ifdim \wd\@tempboxa > 6in
           {\begin{center}
        \parbox{6in}{\footnotesize\baselineskip=12pt Fig.~\thefigure. #1}
            \end{center}}
        \else
             {\begin{center}
             {\footnotesize Fig.~\thefigure. #1}
              \end{center}}
        \fi}

\newcommand{\tcaption}[1]{
        \refstepcounter{table}
        \setbox\@tempboxa = \hbox{\footnotesize Table~\thetable. #1}
        \ifdim \wd\@tempboxa > 6in
           {\begin{center}
        \parbox{6in}{\footnotesize\baselineskip=12pt Table~\thetable. #1}
            \end{center}}
        \else
             {\begin{center}
             {\footnotesize Table~\thetable. #1}
              \end{center}}
        \fi}

\def\@citex[#1]#2{\if@filesw\immediate\write\@auxout
	{\string\citation{#2}}\fi
\def\@citea{}\@cite{\@for\@citeb:=#2\do
	{\@citea\def\@citea{,}\@ifundefined
	{b@\@citeb}{{\bf ?}\@warning
	{Citation `\@citeb' on page \thepage \space undefined}}
	{\csname b@\@citeb\endcsname}}}{#1}}

\newif\if@cghi
\def\cite{\@cghitrue\@ifnextchar [{\@tempswatrue
	\@citex}{\@tempswafalse\@citex[]}}
\def\citelow{\@cghifalse\@ifnextchar [{\@tempswatrue
	\@citex}{\@tempswafalse\@citex[]}}
\def\@cite#1#2{{$\null^{#1}$\if@tempswa\typeout
	{IJCGA warning: optional citation argument 
	ignored: `#2'} \fi}}

 1
 1
 1

\font\ninerm=cmr9

\def\rsep{R_{\rm sep}}
 
\def\xgo{x_\gamma^{OBS}}
 
\def\xg{x_\gamma} 
\def\ETJ{E_T^{jet}}
 
\def\ETAJ{\eta^{jet}} 
\def\ETAB{\bar{\eta}}

\begin{document}

\centerline{\normalsize\bf PHOTOPRODUCTION OF JETS\footnote{UCL/HEP 97-04, 
Presented
at the Ringberg Workshop `New Trends in HERA Physics', Tegernsee,
Gemernay, 25-30 May 1997.}}

\centerline{\footnotesize J. M. BUTTERWORTH}
\baselineskip=13pt
\centerline{\footnotesize\it Department of Physics and Astronomy, University College London, Gower St.}
\baselineskip=12pt
\centerline{\footnotesize\it London, WC1E 6BT, UK}
\centerline{\footnotesize E-mail: jmb@hep.ucl.ac.uk}
\vspace*{0.3cm}
\centerline{\footnotesize On behalf of the H1 and ZEUS collaborations.}

\vspace*{0.9cm}
\abstracts{The status of HERA data on jet photoproduction is
reviewed, and some suggestions and prospects for further
work are given.}
 
\normalsize\baselineskip=15pt
\setcounter{footnote}{0}
\renewcommand{\thefootnote}{\alph{footnote}}
\section{Introduction}

The photoproduction of jets at HERA is proving to be a very fruitful
process in which to study strong interactions. Aspects of QCD which
are being investigated include the partonic structure of both the
proton and the photon, the internal structure of jets, and the
dynamics of jet production. I will omit jet production in association
with prompt photons, charm and rapidity gaps - these reactions are
covered in other contributions.

\section{Starting Simply}

\begin{figure}
\begin{center}
\psfig{figure=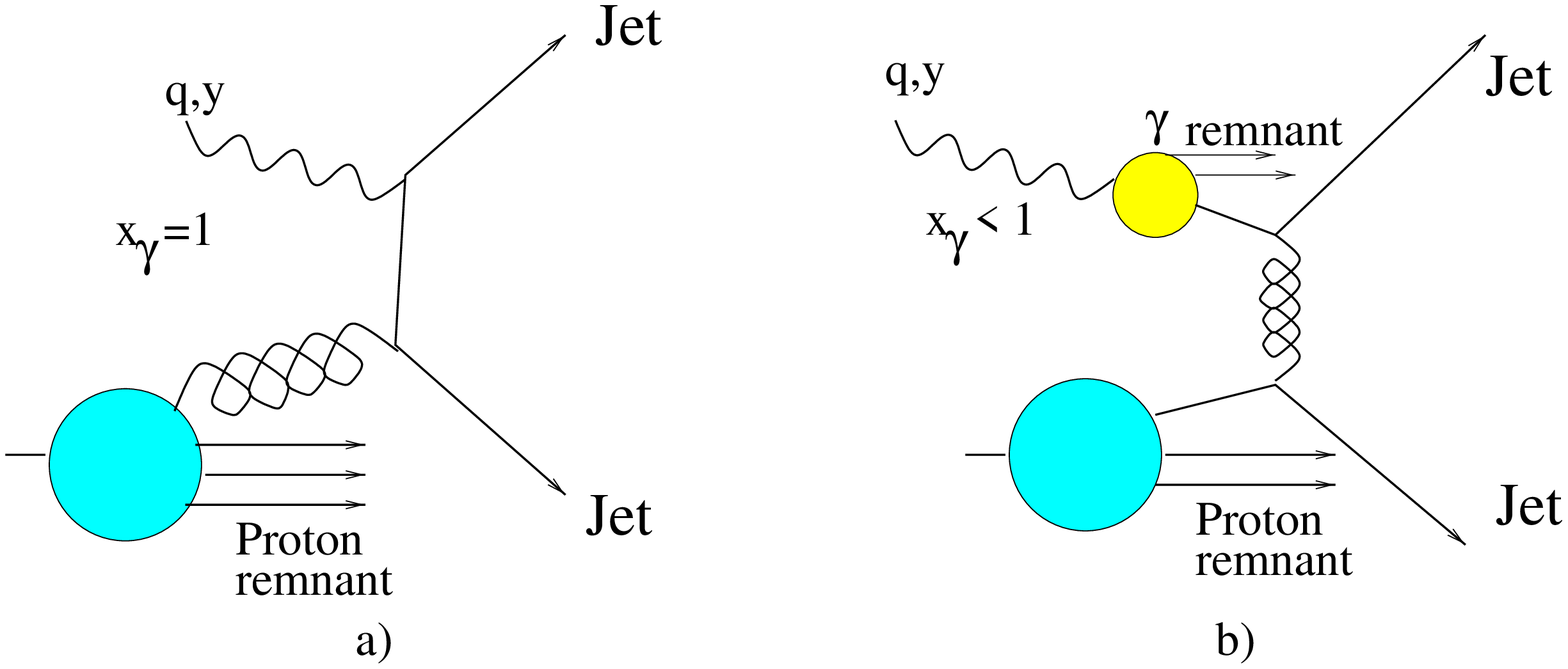,height=5.5cm}
\end{center}
\vspace{-0.2cm}
\fcaption{Leading order diagrams for direct (a) and resolved (b) 
jet photoproduction.}
\label{fig:dirres}
\end{figure}

In leading order QCD, jet photoproduction processes are divided into
two classes: `direct' and `resolved'. Example diagrams for these
processes are shown in Fig.\ref{fig:dirres}. Direct processes are
apparently simple - the photon couples directly into the hard
scattering (via high virtuality quarks), and thus the fraction ($\xg$) 
of its
momentum transferred to the high $E_T$ partons is
one. However, the photon may couple to a $q\bar{q}$ pair with a
relative $p_T$ much less than the $E_T$ of the hard jets. In this case
large logarithms of $E_T/p_T$ enter into calculations of the cross
section and for the lowest $p_T$'s the splitting of the photon is
non-perturbative. These features lead to the introduction of a
partonic structure for the photon, to describe the `cascade' of
partons derived from such low $p_T$ splittings. Reactions involving
partons from this cascade are called resolved photon interactions, and
they enter the jet production cross section at the same order as
direct processes. In this case, $\xg < 1$.

An obvious first question to address at HERA then is - How well does
this leading order language correspond to the actual situation in
experiment? To answer it one must decide carefully what to attempt to
measure. The goal should not be to somehow try to extract `leading
order' cross sections or variables, but to define sensible observables
which have natural interpretations in the LO picture but which are
well defined independently of it.

The first such observable is a jet. At leading order this corresponds
to a parton, but in experiment (as well as in more sophisticated
calculations) it is defined by a jet algorithm. More of this
later. For now, it is enough to remark that jets have well defined
pseudorapidity\footnote{$\eta -\ln \tan {\theta/2}$ where
$\theta$ is the polar angle of the jet w.r.t. the proton direction.}
($\ETAJ$) and transverse energy ($\ETJ$).

In photoproduction events at HERA the positron generally escapes down
the beampipe. This constrains the negative of the four-momentum
squared of the photon ($Q^2$) to be below about 4~GeV$^2$, and the
median value is $Q^2 \approx 10^{-3}$ GeV$^2$. This is {\it not} the
hard scale of the interaction (which is provided by $\ETJ$), and is
more usually referred to as $P^2$ in photon physics. Another important
variable is the inelasticity $y$. This is defined in the same way as
in deep inelastic scattering (DIS), but at low $Q^2$ it reduces to $y
\approx E_\gamma/E_e$.

Using the jets and $y$, the variable $\xgo = \sum_{jets}( E_T^{jet}
e^{-\eta^{jet}} )/ 2yE_e$ is defined. It is the fraction of the
photon's momentum appearing in the high $\ETJ$ jets. In dijet cross
sections the sum runs over the two highest $\ETJ$ events. The
superscript `OBS' is meant to indicate that this is an observable,
unlike the LO $\xg$. However, in a two parton final state, $\xgo$
reduces to the LO variable.

\begin{figure}
\begin{center}
\psfig{figure=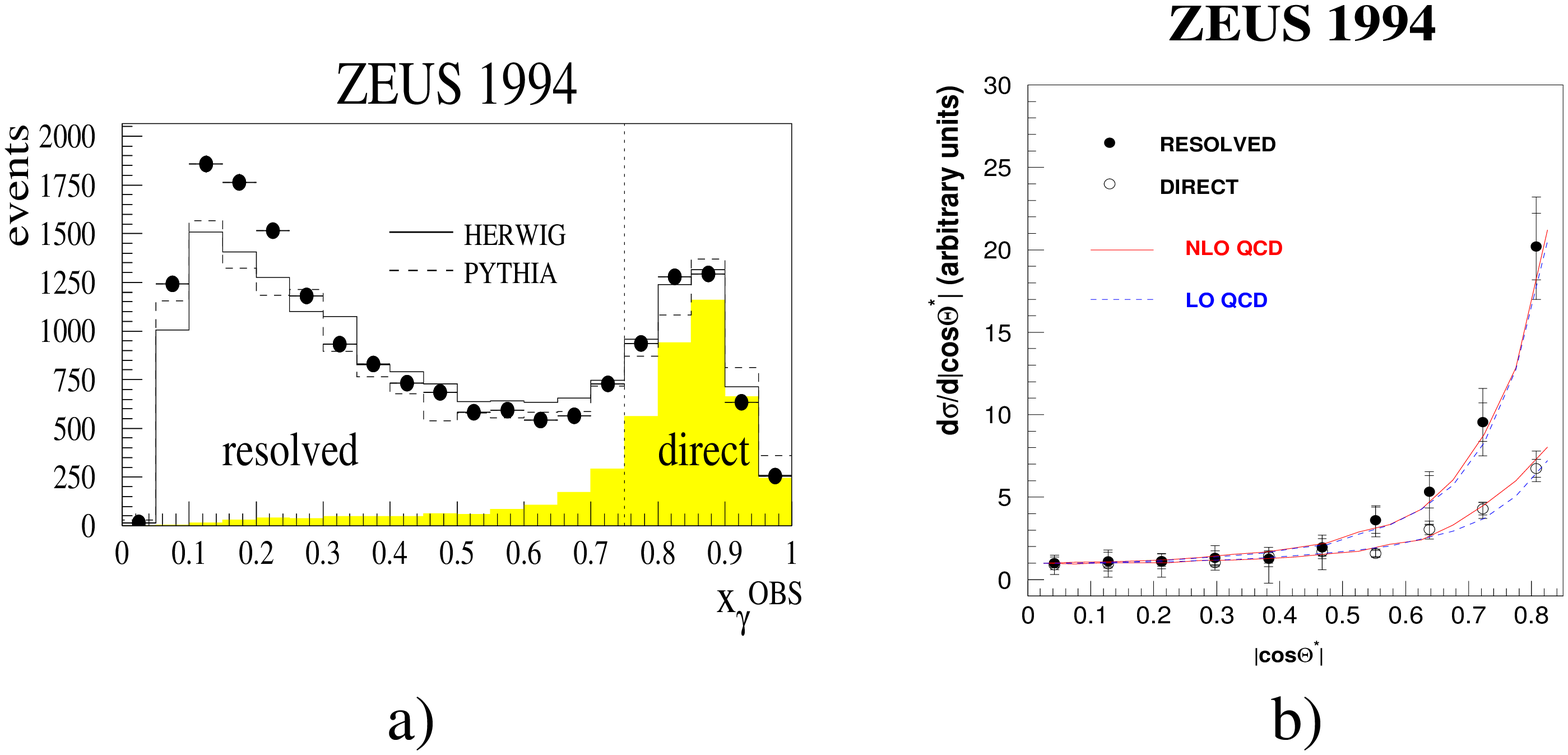,height=7.5cm}
\end{center}
\vspace{-0.5cm}
\fcaption{a) Uncorrected $\xgo$ distribution b) dijet angular distributions.}
\label{fig:xgetc}
\end{figure}

The (uncorrected) $\xgo$ distribution from ZEUS\cite{zeusdij} is shown
in Fig.\ref{fig:xgetc}. There is a clear two-component structure,
with a peak at high values of $\xgo$ and a rising tail to low values,
which is cut off eventually by detector acceptance. Also shown are the
distributions obtained from Monte Carlo (MC) simulations which include
parton showers and hadronisation models as well as the LO
diagrams. The direct MC events lie at high $\xgo$.

Thus the LO QCD picture has passed its first test (at least
qualitatively - there are discrepancies at low $\xgo$ between the data
and the MC). There is a further test which can be made quite
simply. The dominant LO diagrams for direct processes involve fermion
(i.e. quark) exchange whereas those in resolved processes involve
boson (i.e. gluon) exchange (see Fig.\ref{fig:dirres}). This leads to
different predictions for the dijet angular distribution in the
jet-jet centre-of-mass frame. The direct (high $\xgo$) should be
distributed according to $|1-cos{\theta^*}|^{-1}$, whereas the
resolved (low $\xgo$) should be $\approx |1-cos{\theta^*}|^{-2}$. ZEUS
has measured these distributions\cite{zeusang} for dijet invariant
mass above 23~GeV. The results are shown in Fig.\ref{fig:xgetc}. The
data agree well with the predictions.  Also shown are NLO calculations
from Harris and Owens\cite{ho}, which agree well with both the data
and with the LO curves.

\section{Jet Cross Sections} 

So far the simple LO picture of these processes is in pretty good
shape. What else can we learn? As well as being sensitive to the QCD
dynamics, jet cross sections are sensitive to the parton distributions
in the photon and the proton. Thus in principle they can give
information about the quark and gluon distributions inside the photon
and proton. H1 and ZEUS have measured inclusive and dijet cross
sections and the statistics are now becoming sufficiently high for
measurements of multijet jet cross sections to begin.

\subsection{Inclusive Jet Cross Sections} 

Both ZEUS and H1 measure inclusive jet cross sections differential in
$\ETAJ$ and integrated above a given $\ETJ$, and cross sections
differential in $\ETJ$ integrated within a range of $\ETAJ$.  All the
cross sections are $ep$ cross sections integrated within specified $y$
and $Q^2$ ranges.

\begin{figure}
\begin{center}
\psfig{figure=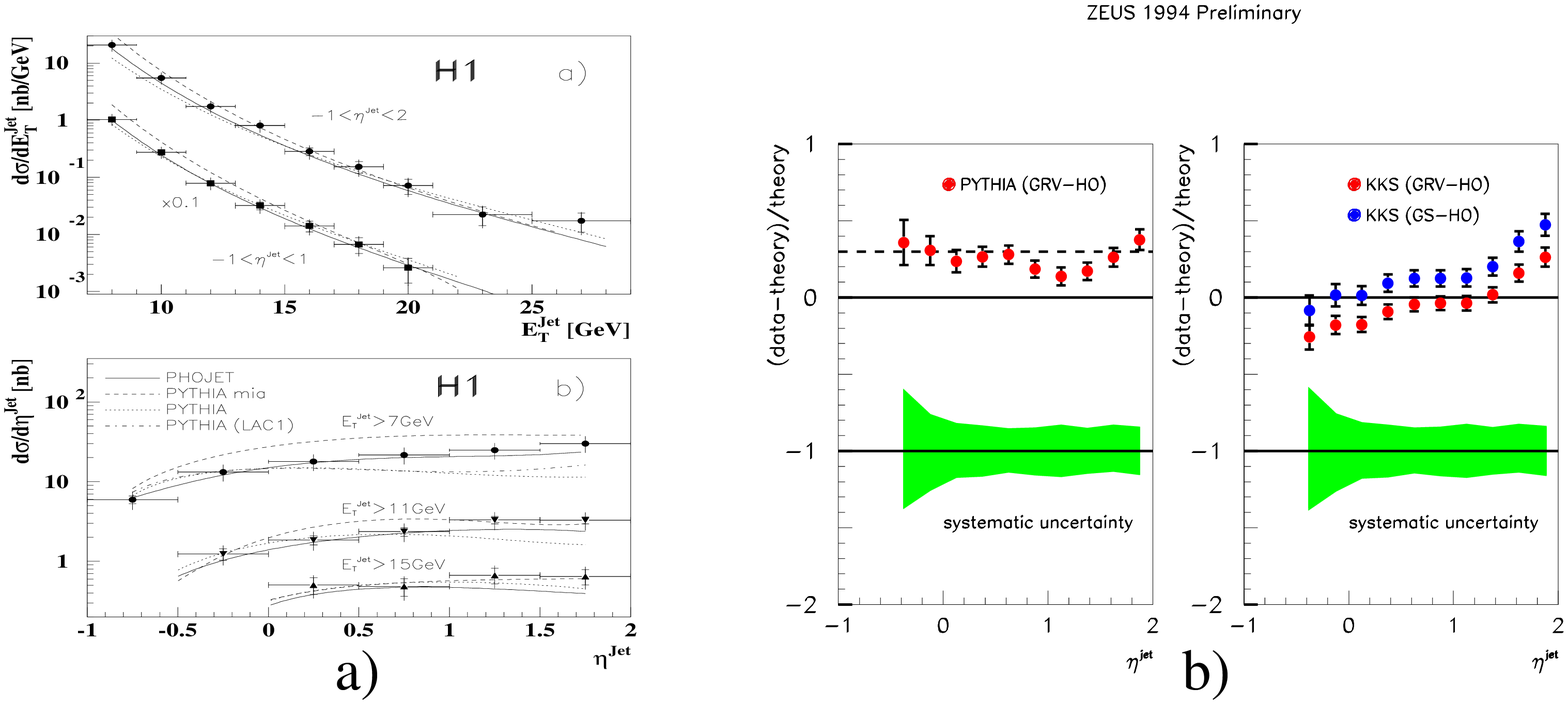,height=10.0cm}
\end{center}
\vspace{-1.5cm}
\fcaption{Inclusive jet cross sections. The H1 data\cite{h1inc} have
a normalisation uncertainty of 26\% which is not shown. The
uncertainty in the ZEUS data arising from the energy scale of the
calorimeter is correlated between points and is indicated by the
shaded band. The ZEUS data show the difference between data and theory
for the cross section $d\sigma/d\ETAJ$ for $\ETJ >
17$~GeV\cite{zeusinc}.}
\label{fig:inc1}
\end{figure}

Some examples are shown in Fig.\ref{fig:inc1}. The H1 data\cite{h1inc}
are compared to the expectations of LO MC simulations. For the
standard PYTHIA there is in general reasonable agreement in the shape
at high $\ETJ$ values. However, at low $\ETJ$ and in the forward
region the data lie above the MC (whether GRV or LAC1 parton
distributions are used for the photon). The other MC models shown
contain multiparton interactions and give higher cross sections - more
of this later. An example of the ZEUS inclusive jet data is also
shown. This time, the difference between data and `theory' is plotted,
where in the first case the theory is PYTHIA again, and in the second
case it is a NLO QCD calculation from Klasen and Kramer\cite{kk}. For
these higher $\ETJ$ values PYTHIA lies below the data over the whole
range. The NLO QCD calculations describe the normalisation of the data
better, but lie below the data in the forward region. The sensitivity
to the parton distribution in the photon is similar in size to the
systematic uncertainties in the measurement.

\subsection{Dijet Cross Sections} 

\begin{figure}
\begin{center}
\hspace{3cm}
\psfig{figure=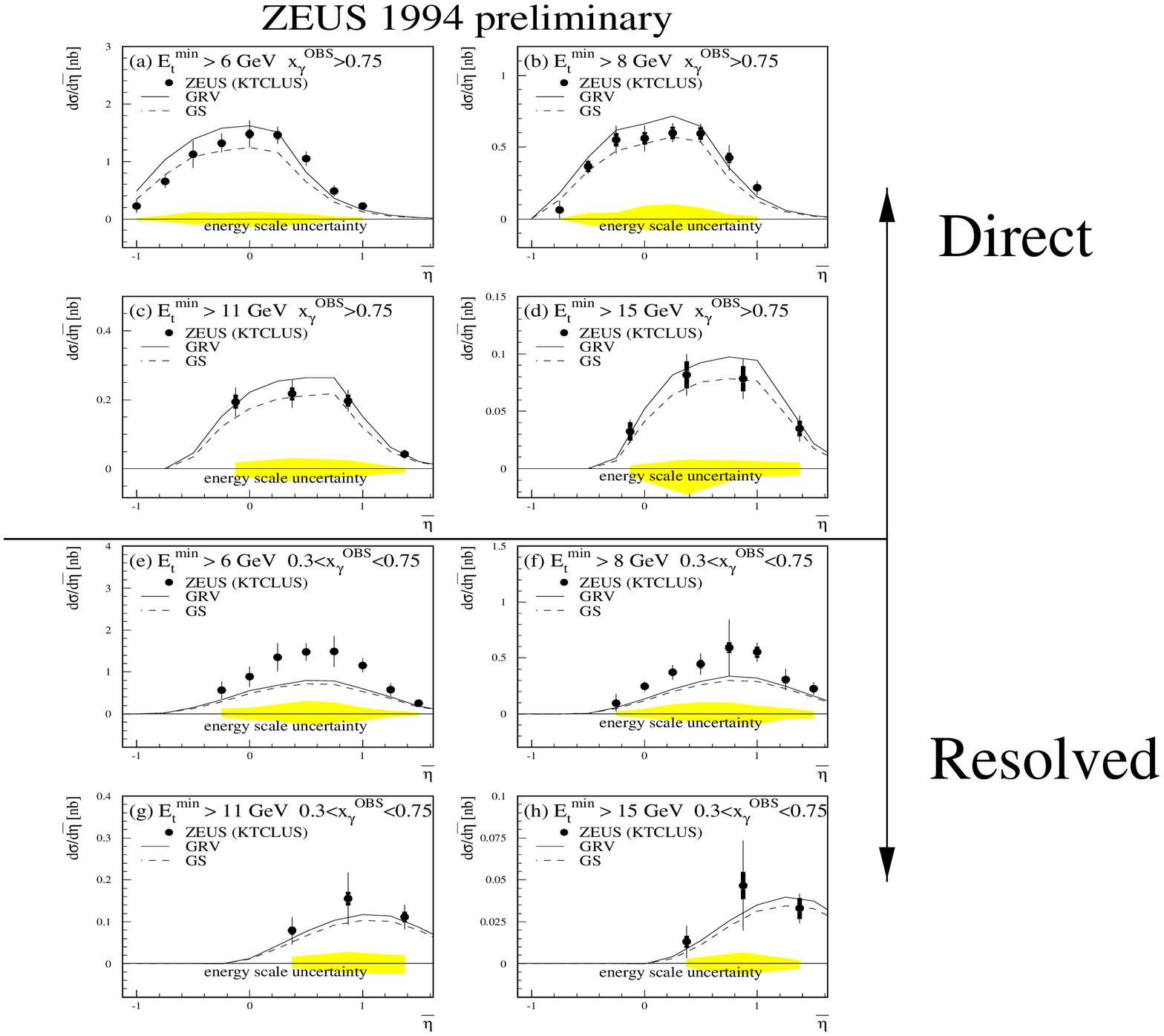,height=15cm}
\end{center}
\vspace{-0.5cm}
\fcaption{Dijet cross sections from ZEUS\cite{zeusdij}.  The
uncertainty in the data arising from the energy scale of the
calorimeter is correlated between points and is indicated by the
shaded band.}
\label{fig:dijets}
\end{figure}

Once two (or more) jets per event are measured in the detector, many
possible cross sections can be measured. The angular distributions
shown earlier are an example, as might be the $\xgo$
distribution. A choice which has been made by ZEUS\cite{zeusdij} is to
measure $d\sigma/d\ETAB$ for both jets above a given $E_T^{jet}$
cut. Here $\ETAB \equiv (\eta_1 + \eta_2)/2$ is the boost of the dijet
system in the lab frame. Rewriting $\xgo \approx \ETJ e^{-\bar{\eta}}
\cosh{\Delta\eta}/y E_e$ shows that for small $|\Delta\eta| \equiv
|\eta_1 - \eta_2|)$ the smallest $x$ values are probed for a given
$\ETJ$. Scanning across $\ETAB$ means scanning across $y x_\gamma$ and
$x_p$. Low $\ETAB$ means high $x_\gamma$ and low $x_p$ (typically
0.005). High $\ETAB$ means low $x_\gamma$ and moderate $x_p$
(typically 0.1). The cross section is measured in two $\xgo$ regions
corresponding to direct ($\xgo > 0.75$) and resolved ($0.3 < \xgo <
0.75$). The data are shown in Fig.\ref{fig:dijets}.

\subsection{Comments on Jet Cross Sections} 

The situation now looks a little less clear. There is reasonable
agreement between data and theory for dijet cross sections at high
$\xgo$, but the calculations are too low at low $\ETJ$ and low $\xgo$.
There is reasonable agreement in backward inclusive jets, but the
calculation is too low in the forward direction, particularly for the
lowest $\ETJ$ values measured. The next section contains some possible
explanations and hints at how these discrepancies might be resolved.

\section{What is a jet?}

The theory we wish to investigate is QCD. Although MC simulations
describe some event properties more successfully by use of
phenomenological models, the best calculations available are at
next-to-leading order in $\alpha_s$ and do not include effects such as
hadronisation. This means that although in nature jets consist of many
(typically more than five) hadrons, in the theory they consist of two
(at most) partons. These partons are the hardest in the event, and
thus can be expected to give a reasonable description of high $\ETJ$
jets. However, the following issues (and maybe others!) must be
addressed before strong conclusions can be drawn from comparisons
between data and theory. None of them are trivial issues, and the
investigation of them should deepen our understanding significantly.

\subsection{Jet Algorithms} 

It is not correct to think of jets as being simply `smeared'
partons. Jets are defined by an algorithm. Are the algorithms the same
in experiment and theory? Experiments have used cone algorithms of
various flavours or more lately a mode of the so-called $k_T$ cluster
algorithm which uses separation in $\eta-\phi$ space as its distance
parameter\cite{ktclus}. So far the calculations have used only a cone
algorithm.

A major source of ambiguity arises in the seed finding (that is, where
do you begin looking for a jet?) and jet merging and/or splitting
(that is, at what point does a fairly hard subjet become an extra jet
in its own right?). The details of how these are performed can have a
large effect on the experiment and a lesser or different one on the
theory. No unique treatment is defined in the famous Snowmass
convention\cite{snowmass}. A particular example of a problem can be
seen by considering two partons or hard particles separated by $\delta
r = \sqrt{\delta\phi^2 + \delta\eta^2} =2$. An experimental cone
algorithm running with jet radius $R = 1$ would typically take one of
these (or the calorimeter cluster caused by it) and draw a circle of
radius one around it. No other particle is inside this radius and so a
stable jet is formed. The second particle will form a second
jet. However, in a theoretical calculation the partons will be merged
(if they have the same $E_T$) because they {\it do} both lie with a
cone of radius one centred on their midpoint.

The parameter $\rsep$ is introduced in the theory to combat this
problem\cite{rsep}.  Partons separated by a distance greater than
$\rsep$ will never be merged by the algorithm. In this way,
calculations attempt to mimic the effects of the seed finding stage of
experimental jet finders. As a bonus, setting $\rsep = R$ makes the
results of the cone algorithm identical (for a three parton final
state) to the $k_T$ version employed by the experiments.

Therefore one approach is to measure and calculate jet cross sections
with various jet algorithms and see how changes in the algorithm
affect the comparison between the two\cite{lutz}. However, the
$\rsep$ method is in fact badly defined for higher order calculations,
and in fact cone algorithms like this are not infrared safe in
four-parton final states\cite{safe}.  

\subsection{Underlying Events} 

Jets are not the only things in an event. How does the rest of the
event affect them? The term `underlying event' is often used loosely
to refer to the activity in real events caused by the fact that,
having provided partons for a hard scattering, the remnants of the
proton and photon do not just go away. The models and language used to
describe them vary widely and depending upon taste or convention
effects such as initial and final state QCD radiation and soft or hard
remnant-remnant interactions may or may not be included. They {\it do
not} however include second photon-proton interactions in a single
bunch crossing - the probability for this leading to significant
activity in an event is negligible at HERA.

\begin{figure}
\begin{center}
\psfig{figure=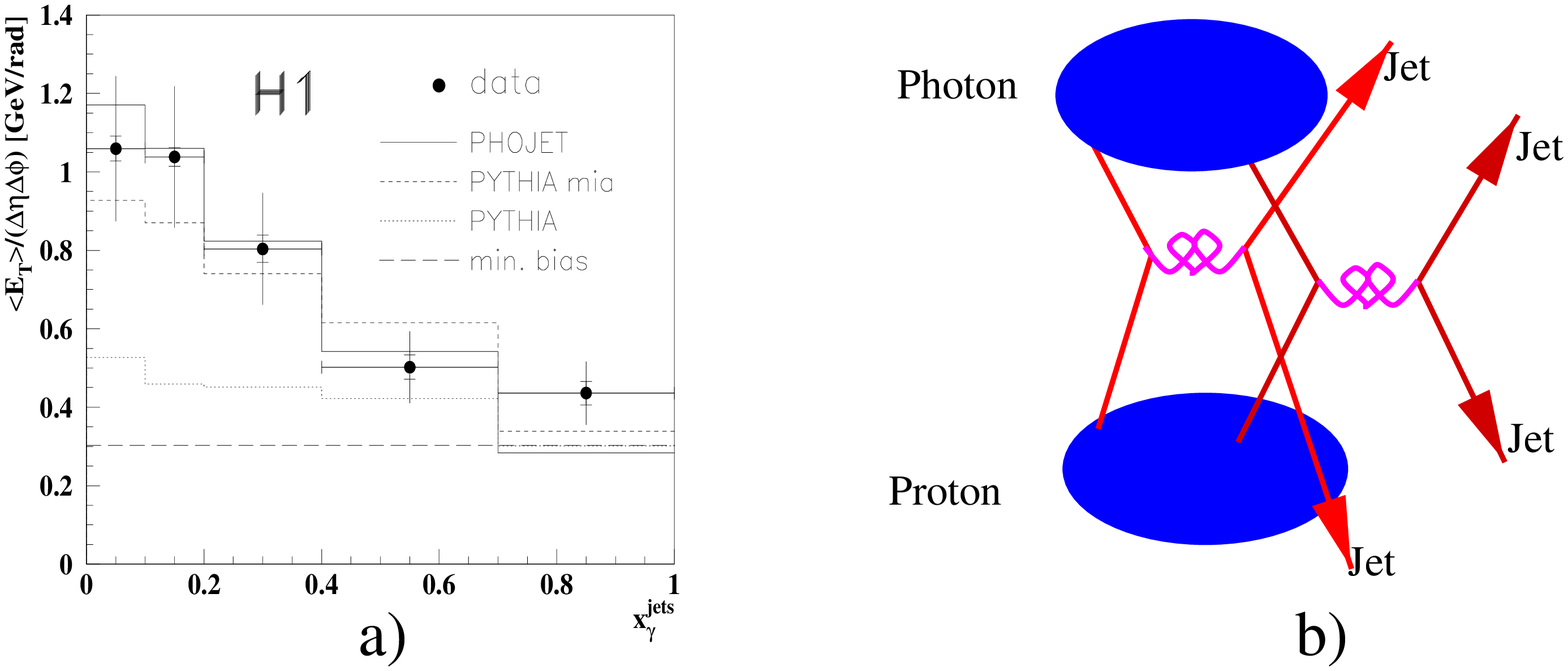,height=8cm}
\end{center}
\vspace{-0.2cm} 
\fcaption{(a) Energy flow outside the jets vs $\xgo$
($\equiv x_\gamma^{jets}$).  (b) Schematic representation of
multiparton interactions.}
\label{fig:etflow}
\end{figure}

That an underlying event exists in photoproduction at HERA is clearly
seen in Fig.\ref{fig:etflow}a. Here H1\cite{h1inc} have plotted the
mean transverse energy {\it outside the jet} per unit of $\eta-\phi$
space, as a function of $\xgo$. First, there is clearly plenty of
transverse energy in the event apart from the hard jets, and secondly
it is correlated to $\xgo$. Given the fact that any transverse energy
outside the jet must contain some of the photon's momentum, it is plain
from the definition of $\xgo$ that such a correlation must
exist. However, the amount of transverse energy involved and strength
of the correlation are remarkable. The implication is that this
underlying event could indeed affect the jets significantly, and that
it cannot be treated as independent of the jets. The size of the
underlying event is correlated to the hard process, and therefore
no technique relying on subtraction of typical `minimum bias' events
will be able to correct for it properly. That such techniques have
occasionally been used at hadron-hadron colliders without large ill
effects seems to be because all the events are at low $x$, whereas the
correlation becomes important in the range above $x \approx 0.1$.

A possible explanation of this type of effect is offered by
multiparton interactions (MI)\cite{mi,bfs}. These are allowed in
eikonal models for extra hard (or sometimes soft) scatters in a single
$\gamma p$ event, as illustrated in Fig.\ref{fig:etflow}b. Such models
are available in PYTHIA\cite{PYTHIA}, HERWIG\cite{HERWIG} and
PHOJET\cite{PHOJET} (but not in NLO QCD!). These models improve the
description of the data - they increase energy flow around the jet
core and in general increase jet cross sections. The price paid is
that when MI are allowed, the energy flow outside the jet becomes very
sensitive to the parton distributions in the photon and proton, and to
$p_T^{\rm min}$, the cutoff for hard scattering\cite{bfs}. This is
because the average number of partonic interactions in a given $\gamma
p$ event goes up with increasing parton density.

\subsection{Jet Shapes} 

\begin{figure}
\begin{center}
\psfig{figure=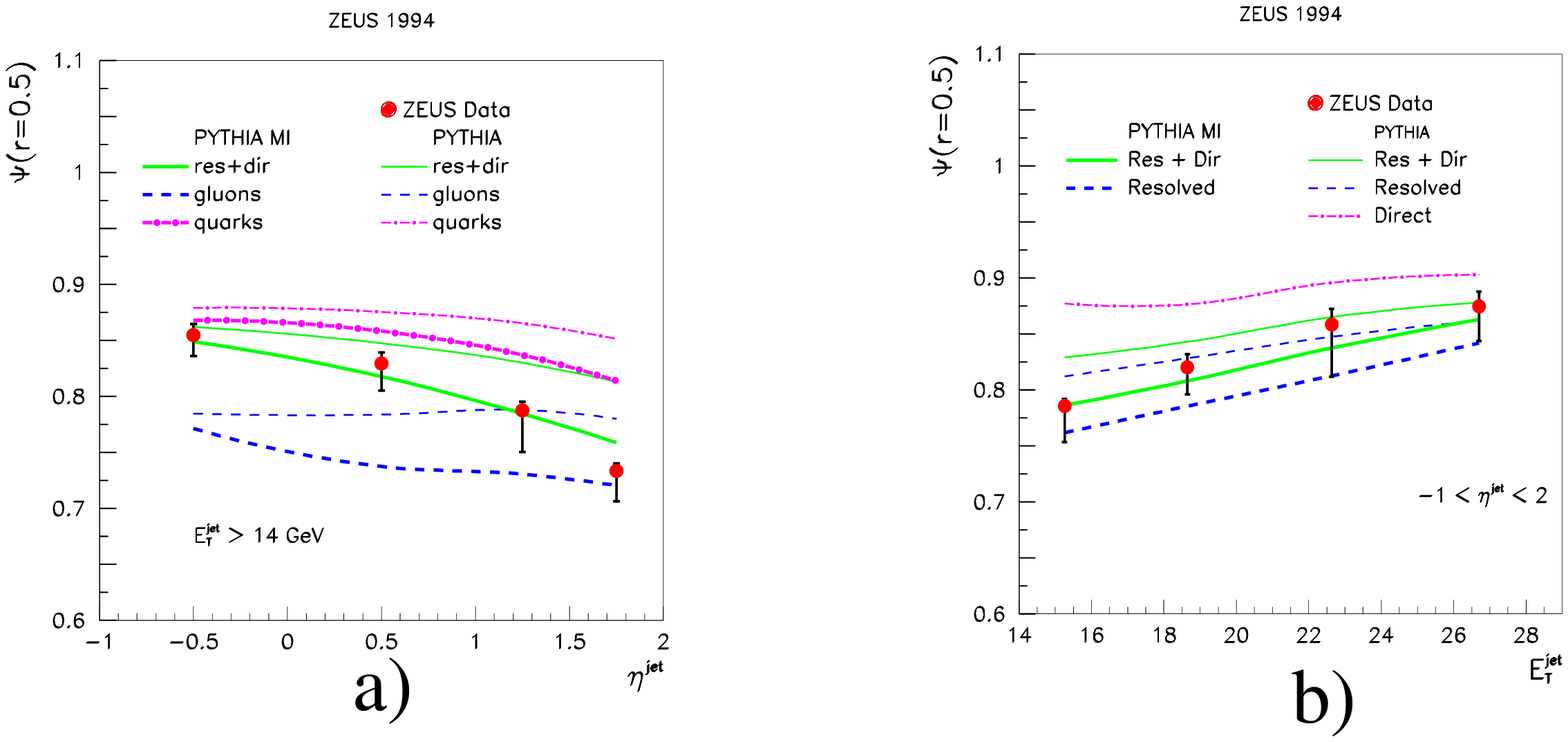,height=9.0cm}
\end{center}
\vspace{-1.5cm}
\fcaption{Jet shapes}
\label{fig:shapes2}
\end{figure}

Jets have internal structure. Does the theory describe this properly?
Measurements of jet shapes provide a useful way of looking at
this. The jet shape $\Psi(r)$ is defined as the average fraction of
the jet's transverse energy lying within cone of radius $r$. It is
defined such that $\psi(R) = 1$. The rate at which it approaches unity
as $r$ approaches $R$ is a measure of how collimated the jet is.  ZEUS
have measured jet shapes\cite{shapes} for a sample inclusive jets with
$E_T^{jet} > 14$ GeV. An example of the data is shown in
Fig.\ref{fig:shapes2}.  In Fig.\ref{fig:shapes2}a the fraction of 
$\ETJ$ contained within a sub-cone of radius 0.5 is plotted as a
function of $\ETAJ$. The data show that forward jets are broader; that
is, the fraction of $\ETJ$ within the inner cone decreases as $\ETAJ$
increases. Also shown are several curves from PYTHIA. The two
continuous lines show the prediction of PYTHIA with (thicker) and
without (thinner) MI. Clearly, MI once more improve the description of
the data. Also shown separately are the shapes for those PYTHIA jets
which are initiated by a gluon or a quark at LO. Gluon jets are
broader in the MC, as expected from the fact that their larger colour
charge leads to more QCD radiation. The jets in the rear direction
look like quark jets, whereas the behaviour is gluon like in the
forward direction. The transition between the two is reproduced in the
MC, where it arises from the transition from dominantly direct to
dominantly resolved processes as $\ETAJ$ increases. In
Fig.\ref{fig:shapes2}b, the same quantity is plotted, but now as a
function of $\ETJ$. Jets get narrower as $\ETJ$ increases. In
addition, in the MC the effect of multiparton interactions decreases.
Comparisons to QCD calculations are given in Klasen's presentation.


\section{Outlook and Conclusion} 

A theme in this area of physics is the benefit (and difficulty) of
making general comparisons. There has been significant progress in our
understanding of how to compare data and theory. The choice of jet
algorithm has been discussed in this context and an attractive
solution to the problems involved seems to be offered by $k_T$
algorithm, which by virtue of being a cluster rather than cone
algorithm avoids in a natural manner all seed finding and jet merging
ambiguities (and hence the need for $\rsep$), but which in the chosen
mode preserves the attractive features of cone algorithms in hadronic
physics.

Another point to bear in mind is that we need to choose `theory
friendly' cross sections. In particular, demanding two jets above a
certain $E_T$ leads to divergences in NLO QCD which are responsible
for discrepancies between theoretical calculations\cite{ho,kk}. The
problem is discussed in Klasen's contribution. Recent preliminary ZEUS
data\cite{mhayes} avoid this problem by symmetrising dijet cross
sections and cutting on the highest $\ETJ$ whilst applying a different
and lower threshold to the second jet, also shown by Klasen.

To make comparisons at lower $\ETJ$ with confidence, a better
understanding of the `underlying event' is required. Multiparton
interactions are a hot topic, but have many free parameters and are
based upon simplifying assumptions which may not be justified. The
eventual answer is surely to measure and constrain them with data. A
promising approach is to study multijet events, and the current
statistics are now allowing this to begin at HERA\cite{esther}.

Leaving aside the theory for a moment, comparisons between data from
different experiments would be greatly enhanced by the adoption of a
standard set of cuts and jet algorithms for a minimal subset of jet
cross sections, where this is practical. The comparison of jet physics
between photoproduction and DIS is another subject which is gaining
momentum, both in jet shapes and cross sections.

\begin{figure}
\begin{center}
\psfig{figure=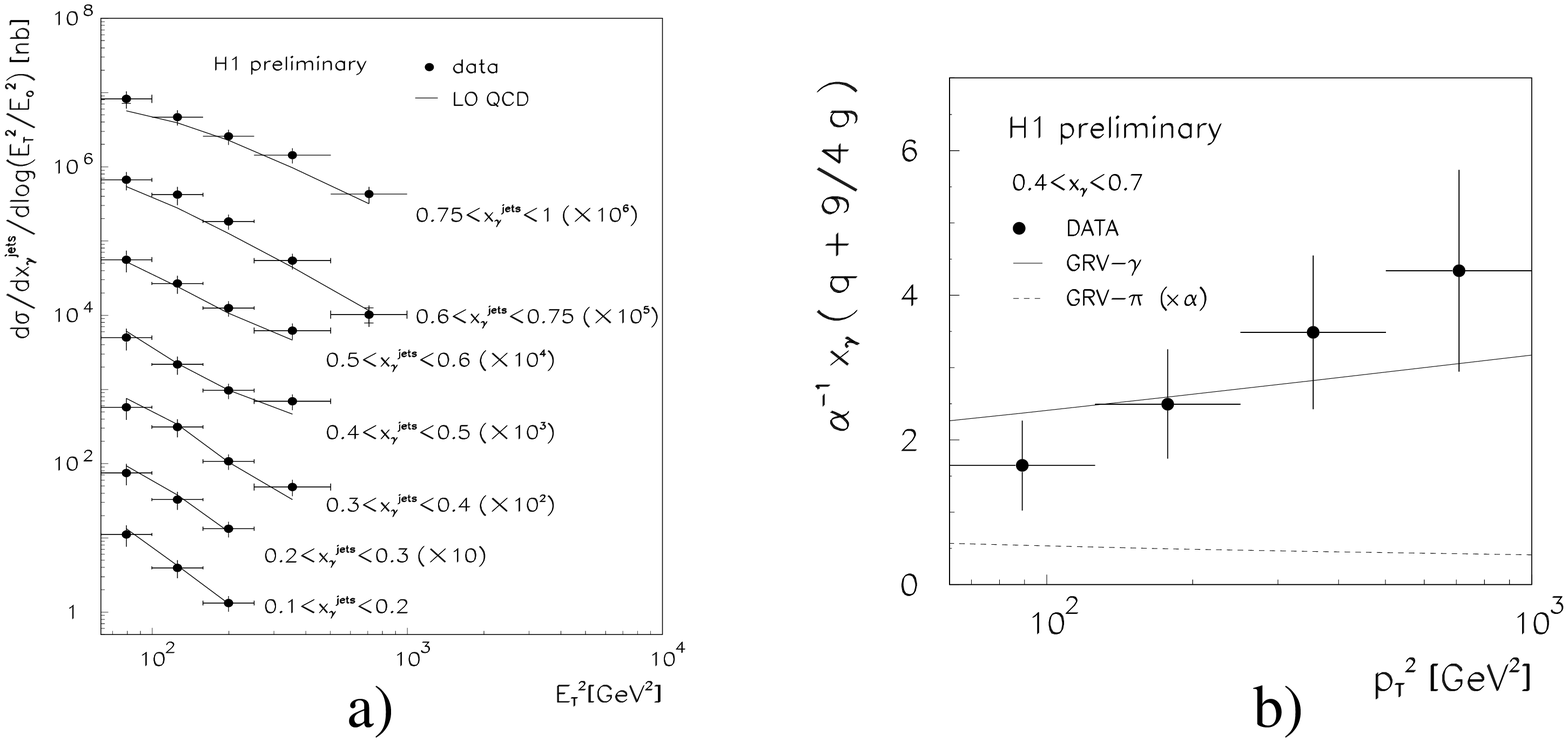,height=8.5cm}
\end{center}
\vspace{-0.5cm}
\fcaption{(a) Dijet cross sections, (b) Effective parton distribution
in the photon.}
\label{fig:eff}
\end{figure}

Photoproduction at HERA and $\gamma\gamma$ collisions at $e^+e^-$
experiments probe the photon in complementary ways. Global fits to
both jet data and $F_2^\gamma$ data similar to those carried out for
the proton will probably place the strongest constraints on the
structure of the photon in the end. However, general physics messages
can been drawn from the data in other ways. A promising approach here
is to study the effective parton distributions in the
photon\cite{combridge}. In resolved photoproduction, the most
important matrix elements have the same angular dependence and
contribute to the cross section basically according to the colour
factors involved.
This fact allows the dijet cross section to be written in terms of
single effective matrix element and an effective parton distribution:
\begin{displaymath}
\frac{d^4\sigma}{dy dx_\gamma dx_p d\cos\theta^*} =
\frac{1}{32\pi s_{ep}} . \frac{f_{\gamma/e}}{y} .
\frac{f^\gamma_{eff}}{x_\gamma} .
\frac{f^p_{eff}}{x_p} .
|M_{eff}|^2
\end{displaymath}
where $f_{\rm eff}(x,p_T^2) = \sum
q(x,p_T^2)+\bar{q}(x,p_T^2)+\frac{9}{4}g(x,p_T^2)$.  H1\cite{h1eff}
have measured the dijet cross section shown in Fig.\ref{fig:eff}a,
and extracted $f^\gamma_{\rm eff}(x,p_T^2)$ as shown in
Fig.\ref{fig:eff}b. There is a positive scaling violation, as seen in
$F_2^\gamma$. This is characteristic of the $\gamma \rightarrow
q\bar{q}$ splitting probability, and is not present in purely hadronic
structure functions.

Finally, a nice instance of feedback between the two types of
experiments was recently shown in OPAL data\cite{jan}. In 1993, ZEUS
results showed that the data are better described by MC simulations if
the intrinsic $k_T$ in the photon is increased\cite{zeusrem}. This is
a way of faking the effect of the fact that in the $\gamma \rightarrow
q\bar{q}$ splitting the quarks can have significant relative $p_T$.
Following this, OPAL have used a similar trick in $e\gamma$ collisions
to achieve a significantly better agreement in the spectrum of
transverse energy outside the plane defined by the beam and the
scattered electron, as well as in the dijet rates for these events.

I hope some flavour of the rapid and continuing progress in field has
been conveyed. I am confident that there is a great deal more physics
to come from studying jet photoproduction.

\section{Acknowledgements}
My thanks go to the many people on H1 and ZEUS who provided plots.  I
would also like to thank the DESY directorate and MPI, as well as the
organisers and participants of what was an immensely stimulating and
enjoyable meeting.

\end{document}

\end{document}